# Cylindrical cavity expansion analysis under partially drained conditions for normalisation of excess water pressure in CPTU

Dr He **Yang**
E-mail: cnhy@leeds.ac.uk
School of Civil Engineering, University of Leeds, Leeds, LS2 9JT, UK

Professor Pei-Zhi **Zhuang**
E-mail: zhuangpeizhi@sdu.edu.cn
School of Qilu Transportation, Shandong University, Jinan, 250002, China

Professor Hai-Sui **Yu**, FREng
E-mail: H.Yu@leeds.ac.uk
School of Civil Engineering, University of Leeds, Leeds, LS2 9JT, UK

Dr Pin-Qiang **Mo**, Associate Professor
Email: pinqiang.mo@cumt.edu.cn
School of Mechanics and Civil Engineering, China University of Mining and Technology, Xuzhou, 221116, China

Dr Yue **Ma**
E-mail: mayue_1990cn@hotmail.com
School of Civil Engineering, University of Leeds, Leeds, LS2 9JT, UK

Dr Xiaohui **Chen**, Associate Professor
E-mail: X.Chen@leeds.ac.uk
School of Civil Engineering, University of Leeds, Leeds, LS2 9JT, UK

Professor Fernando **Schnaid**
E-mail: fschnaid@gmail.com
School of Civil Engineering, Universidade Federal do Rio Grande do Sul, Brazil

**Abstract:**

Cone tip resistance and excess water pressure (EWP) measured by piezocone penetration tests (CPTU) may be significantly affected by the partially drained effect in soils with intermediate permeability. To capture this effect, the paper proposes a straightforward, hydro-mechanical coupling solution for cylindrical cavity expansion under partially drained conditions. The mechanical behaviour of soils is modelled using the elastoplastic Tresca model and water flow within porous soils is assumed to obey Darcy's law. Two partial differential equations (PDEs) are established in the elastic and plastic zones, respectively, transforming the cavity expansion analysis into a typical Stefan problem with dynamic boundary conditions (i.e. a moving boundary at the elastoplastic interface). An approximate solution for the PDEs is derived by leveraging the variable transformation method. Based on the new solution, a novel normalised penetration rate is defined considering the rigidity index of soils, with which a unique backbone curve for CPTU is found. Finally, the backbone curve is compared with a database comprising 109 in-situ experimental tests, 101 centrifuge modelling tests, and numerical simulation results. The proposed solution may provide a useful theoretical tool for interpreting the consolidation coefficient of fine-grained soils from the penetration stage of multi-rate CPTU, which can enhance the interpretation reliability for CPTU dissipation tests.

## 1. Introduction

The piezocone penetration test (CPTU) is an efficient and economic in-situ test method to interpret soil properties such as strength, stiffness, and permeability from the measured cone tip resistance and water pressure (Robertson and Campanella 1983a; Robertson and Campanella 1983b; Ghafghazi and Shuttle 2008; Lunne 2012; Mayne and Peuchen 2022). For a standard penetration rate (i.e. 20mm/s), it is normally assumed that CPTU is conducted under fully undrained conditions in low permeable clays and fully drained conditions in high permeable sands. However, partially drained conditions may prevail for CPTU in intermediately permeable soils, such as silty soils and some silty/clayed tailings (Abu-Farsakh et al. 2003; DeJong and Randolph 2012; Sheng et al. 2014; Russell et al. 2023). Besides, various penetration rates in a given soil can be another reason for the partially drained effect, namely the rate effect (Silva et al. 2006; Schneider et al. 2007; Kim et al. 2008; DeJong and Randolph 2012).

Since the last two decades, the partially drained effect in CPTU has been investigated with various methods, including the finite element method (FEM) (Liyanapathirana 2009; Obrzud et al. 2011; Yi et al. 2012; Sheng et al. 2014; Liu et al. 2024), material point method (MPM) (Ceccato et al. 2016; Ceccato and Simonini 2017), particle finite element method (Monforte et al. 2021), in-situ tests (Bedin 2006; Kim et al. 2008; Klahold 2013; Suzuki 2015; Sosnoski 2016), centrifuge model tests (House et al. 2001; Randolph and Hope 2004; Schneider et al. 2007; Jaeger et al. 2010), and cavity expansion method (Silva et al. 2006; Dienstmann et al. 2018; Mo et al. 2020; Reid and Smith 2021; Zhou et al. 2021b; Mafra and Dienstmann 2022; Russell et al. 2023). These studies provided important evidence that the partially drained effect has an essential influence on the evolution of measured excess water pressure (EWP) and cone tip resistance.

In the above studies with various methods, the partially drained effect in CPTU was normally estimated by a normalised penetration rate, $\overline{V}_0$, defined as (Randolph and

Hope 2004)

$$\bar{V}_0=\frac{V_{\text{cptu}}D}{c_{\text{h}}} \text{ or } \frac{V_{\text{cptu}}D}{c_{\text{v}}} \tag{1}$$

where $V_{\text{cptu}}$=penetration rate; $D$=piezocone diameter; $c_{\text{h}}$ and $c_{\text{v}}$ are the horizontal and vertical coefficients of consolidation, respectively. It is recognised that CPTU is conducted under partially drained conditions for $\bar{V}_0$=0.01-30, undrained conditions for $\bar{V}_0$>30, and drained conditions for $\bar{V}_0$<0.01 (Randolph and Hope 2004). However, previous studies indicate the need to improve the normalisation process by incorporating the influence of the rigidity index of soils, $I_{\text{r}}$, in that:

(a) Normalised EWP and cone tip resistance by $\bar{V}_0$ (i.e. the backbone curves in CPTU) were found to be significantly affected by $I_{\text{r}}$ (Dienstmann et al. 2017; Dienstmann et al. 2018; Mafra and Dienstmann 2022).

(b) Since Teh and Houlsby (1991) proposed a method of interpreting consolidation coefficient from CPTU dissipation tests (i.e. holding test that records water pressure dissipation with time at a certain depth), the effect of $I_{\text{r}}$ has been widely considered in the interpretation of those dissipation tests (Chai et al. 2012; Cai et al. 2015; Chai et al. 2016; Zhang et al. 2022; Ecemis et al. 2023). Consideration of $I_{\text{r}}$ ought to be equally important for the interpretation of CPTU in the penetration process.

Cavity expansion method, which investigates the changes in stresses and displacements around an expanding cylindrical or a spherical cavity, has proved to be a simple and efficient tool in geotechnical engineering (Yu 2000). While previous cavity expansion solutions were mainly restricted to fully drained and/or undrained conditions (Yu and Houlsby 1991; Collins et al. 1992; Collins and Yu 1996; Chen and Abousleiman 2013; Chen and Wang 2022; Chen and Abousleiman 2023; Yang et al. 2023; Yang et al.

2024), only a few studies developed cavity expansion solutions for the analyses of partially drained effect in CPTU (Silva et al. 2006; Dienstmann et al. 2017; Mo et al. 2020; Zhou et al. 2021a; Mafra and Dienstmann 2022). For example, Silva et al. (2006) and Zhou et al. (2021b) idealised the CPTU penetration process into a rate-dependent cylindrical cavity expansion model in Cam Clay soils, where the rate effect is emphasised for cavity expansion in Boston blue clay and Kaolin clay. Dienstmann et al. (2017) and Mafra and Dienstmann (2022) developed cavity expansion solutions in a poro-elastic-plastic Drucker-Prager (DP) medium by the time-stepping approach (i.e. discretizing the loading process into finite time intervals). In their studies the EWP distribution was predefined to simplify the hydro-mechanical coupling equations, which should be carefully selected in order to improve solution accuracy. Mo et al. (2020) developed a cavity-expansion-based method that linked the partially drained conditions with fully drained and undrained conditions by empirical relationships. Most recently, Russell et al. (2023) presented a time-stepping cavity expansion solution under partially drained conditions with the modified Cam-Clay model (MCC). Seven first-order ordinary differential equations should be solved numerically and iterations are required for the stress and EWP continuity at the elastoplastic boundary. Overall, only time-stepping cavity expansion solutions are available for the analyses of CPTU under partially drained conditions, which may involve cost of calculation time and indirect physical explanations (e.g. obstacles for dimensional analyses). Therefore, it is essential to develop such an elasto-plastic solution in a fairly straightforward form to provide more efficient computation and more insightful understanding of inherent mechanism, for cavity expansion and CPTU under partially drained conditions.

Considering the contributions from these studies, this paper proposes a straightforward-form solution for cylindrical cavity expansion in intermediately permeable soils under partially drained conditions. The mechanical behaviour is modelled by the elastoplastic Tresca model while the hydraulic behaviour obeys Darcy's law. Then two partial differential equations (PDEs) are established in the elastic and plastic zones, respectively, and are solved by the variable transformation method. With the superiority of the present solution in dimensional analyses, a new normalised penetration rate is proposed by considering the influence of rigidity index. It is found that a unique backbone curve exists with the new normalised rate, and the curve matches well with a database of numerical and experimental results. Finally, a method is presented to estimate the consolidation coefficient of soils from multi-rate CPTU (penetration stage), which can provide guidance and redundancy in the interpretation of CPTU dissipation tests.

## 2. Problem Definition and Assumptions

### *2.1. Cavity expansion model for the CPTU analysis*

Modelling penetration problems with simplified geometries and boundary conditions can help understand the mechanisms of the CPTU penetration process. Following former studies on CPTU under partially drained conditions (Silva et al. 2006; Dienstmann et al. 2018; Zhou et al. 2021b; Mafra and Dienstmann 2022), the changes in radial stresses, deformation, and EWP around the cone tip during the penetration process are modelled by a time-dependent expansion of a cylindrical cavity, as shown in Figure 1. The rates of CPTU penetration and radial cavity expansion are denoted as $V_{\mathrm{cptu}}$ and $V_a$, respectively, and they can be related by $V_a = V_{\mathrm{cptu}} \tan(\vartheta/2)$ ($\vartheta$ is the cone tip angle). At the beginning (i.e. initial state before penetration/cavity expansion), total horizontal and vertical stresses ($\sigma_{\mathrm{h0}}$, $\sigma_{\mathrm{v0}}$) act throughout the soil that is homogenous, isotropic, and fully saturated. When the piezocone moves through section XX (see Figure 1), the inner cavity radius will expand from $a_0$ to $a$ and the inner cavity pressure will increase from $\sigma_{\mathrm{h0}}$ to $\sigma_a$. For convenience, the cavity expansion problem is regarded as a plane strain problem (infinitely long in the vertical direction) and a cylindrical coordinate $(r, \theta, z)$ with the origin at the cavity centre is used for performing the theoretical analysis.

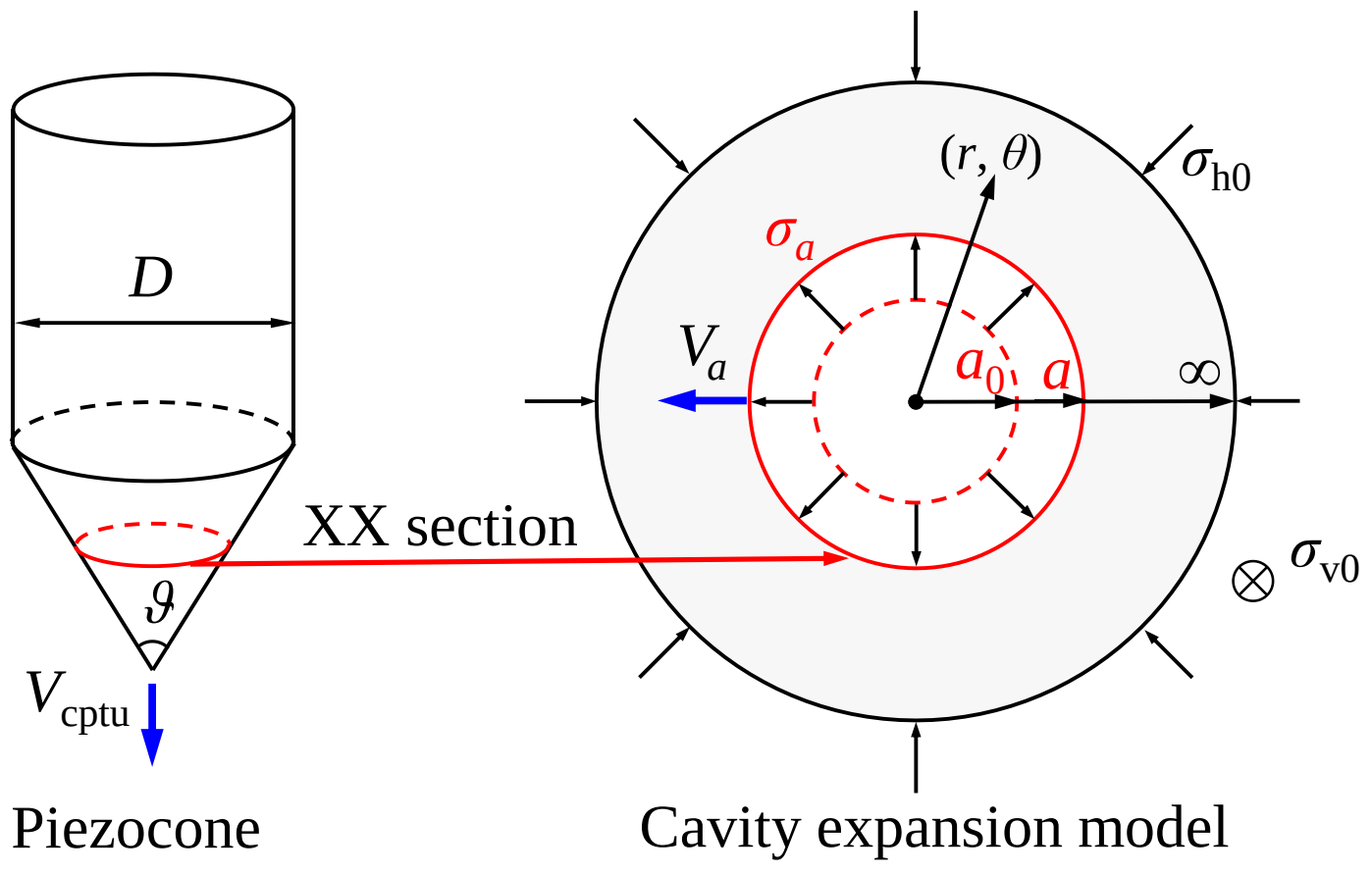

**Figure 1 Relation between the piezocone penetration and cavity expansion**

Alt words: The figure shows a piezocone with the diameter of $D$ is modelled by a cavity expansion mode. The inner cavity radius is initially $a_0$ and then becomes $a$ with loading.

### *2.2. Equilibrium equations*

Ignoring body forces and taking compression as positive, the radial equilibrium equation during cavity expansions can be written in two typical forms as

$$\frac{\partial \sigma_r}{\partial r}+\frac{\sigma_r-\sigma_\theta}{r}=0 \tag{2}$$

$$\frac{\partial \sigma'_r}{\partial r}+\frac{\sigma'_r-\sigma'_\theta}{r}+\frac{\partial U}{\partial r}=0 \tag{3}$$

in which $\sigma_r$ and $\sigma_\theta$ denote the total radial and circumferential stresses, respectively; $\sigma'_r$ and $\sigma'_\theta$ represent the effective radial and circumferential stresses, respectively; $r$ is the current radial position of a soil particle whose initial value is $r_0$; $U$ means EWP relative to the initial ambient water pressure $u_{w0}$.

### *2.3. Stress-strain relationship*

In order to obtain a closed-form solution for the cavity expansion analysis under partially drained conditions, the soil is modelled as a perfectly elastoplastic Tresca model following the well-known work of Teh and Houlsby (1991) (dissipation stage of CPTU). The elastic behaviour of soils is assumed to follow the small strain theory and Hooke's law, as

$$\begin{cases} \varepsilon_r^{\mathrm{e}} = -\dfrac{\partial u_r}{\partial r} = \dfrac{1-\mu}{2G}\left[\left(\sigma_r' - \sigma_{\mathrm{h}0}'\right) - \dfrac{\mu}{1-\mu}\left(\sigma_\theta' - \sigma_{\mathrm{h}0}'\right)\right] \\ \varepsilon_\theta^{\mathrm{e}} = -\dfrac{u_r}{r} = \dfrac{1-\mu}{2G}\left[\left(\sigma_\theta' - \sigma_{\mathrm{h}0}'\right) - \dfrac{\mu}{1-\mu}\left(\sigma_r' - \sigma_{\mathrm{h}0}'\right)\right] \end{cases} \tag{4}$$

where $\varepsilon_r^{\mathrm{e}}$ and $\varepsilon_\theta^{\mathrm{e}}$ are the elastic radial and circumferential strains, respectively; $u_r = r - r_0$ denotes the radial displacement of the soil; $G$ and $\mu$ are the shear modulus and Poisson's ratio; $\sigma_{\mathrm{h}0}' = \sigma_{\mathrm{h}0} - u_{\mathrm{w}0}$ denotes the initial horizontal effective stress.

Once yielding occurs, the soil obeys the Tresca yielding criteria with an associated flow rule, which can be defined as

$$\sigma_r' - \sigma_\theta' = 2s_{\mathrm{u}} \tag{5}$$

$$\frac{\varepsilon_r^{\mathrm{p}}}{\varepsilon_\theta^{\mathrm{p}}} = \frac{\varepsilon_r - \varepsilon_r^{\mathrm{e}}}{\varepsilon_\theta - \varepsilon_\theta^{\mathrm{e}}} = -1 \tag{6}$$

where $s_{\mathrm{u}}$ is the shear strength of soils; $\varepsilon_r^{\mathrm{p}}$ and $\varepsilon_\theta^{\mathrm{p}}$ are the plastic radial and circumferential strains, respectively; $\varepsilon_r$ and $\varepsilon_\theta$ are the total radial and circumferential strains, respectively; the vertical stress is taken as the intermediate principal stress in Equation (5) (Carter et al. 1979; Carter et al. 1986; Yu and Houlsby 1991). Obviously, modelling the soil as a Tresca material is a great simplification in the cavity expansion analysis. Advantages and limitations will be further discussed in the later part of this paper and we will show how the simple soil model can be used for the EWP normalisation for CPTU under partially drained conditions.

### *2.4. Water flow equation*

It is assumed that soil particles and water in soils are incompressible during the penetration process and water flow in the porous soils obeys Darcy's law. For the mass balance of water in saturated soils, the volume change is related to the net water flow, thereby giving (Russell et al. 2023):

$$\frac{\partial \varepsilon_{\mathrm{v}}}{\partial t} + \dot{r}\frac{\partial \varepsilon_{\mathrm{v}}}{\partial r} = -\frac{k}{\gamma_{\mathrm{w}}}\left(\frac{\partial^2 U}{\partial r^2} + \frac{1}{r}\frac{\partial U}{\partial r}\right) \tag{7}$$

where $\varepsilon_v$ is the total volumetric strain; *t* denotes time and *t*=0 for the initial state before loading; $k$ represents the permeability coefficient of soil (soil anisotropy is not considered); $\gamma_w$ represents the specific gravity of water. It is necessary to mention that Eq. (7) is shown in the large-strain form for CPTU analysis. However, the convection term, $\dot{r}\frac{\partial \varepsilon_v}{\partial r}$, is neglected, enabling the solution to be derived in an analytical-form. Yu and Carter (2002) demonstrated that the error induced from this simplification is less than 1% for Tresca model, and rigorous FEM simulations (in Section 6.1) also shows that this convection term is negligible.

### *2.5. Boundary conditions and initial conditions*

At *t*=0, the initial conditions for the defined problem can be written as

$$\begin{cases} \sigma_r = \sigma_\theta = \sigma_{h0} \\ U = 0 \end{cases} \quad \text{for } r \geq a_0 \text{ and } t=0 \tag{8}$$

The boundary conditions at $r = \infty$ and $r = a$ can be summarised as

$$\begin{cases} \sigma_r = \sigma_\theta = \sigma_{h0} \\ U = 0 \end{cases} \quad \text{for } r = \infty \text{ and } t \geq 0 \tag{9}$$

$$\partial U / \partial r = 0 \qquad \text{for } r = a \text{ and } t \geq 0 \tag{10}$$

where Equation (10) means the surface of a piezocone is impermeable (Soderberg 1962; Randolph and Wroth 1979).

## 3. Solution in the elastic zone

As the cavity expands, the surrounding soil is initially in the purely elastic stage. Then a plastic region is formed stemming from the inner cavity wall to $\rho$, where $\rho$ denotes the current radius of the elastoplastic interface. This section shows the analytical solution for the effective stresses, EWP, and displacements in the elastic zone ($r > \rho$).

The compatibility equation in terms of stress components can be obtained by Equation (4) as

$$\frac{\partial}{\partial r}\left[-\mu\sigma_r' + (1-\mu)\sigma_\theta'\right] + \frac{\sigma_\theta' - \sigma_r'}{r} = 0 \tag{11}$$

Combining Equations (3) and (11), the effective stresses and EWP satisfy Equation (12):

$$\frac{\partial}{\partial r}\left[(1-\mu)(\sigma_r' + \sigma_\theta') + U\right] = 0 \tag{12}$$

Then the integral form of Equation (12) over the interval $[\infty,\ r]$ gives

$$(1-\mu)(\sigma_r' + \sigma_\theta') + U = 2(1-\mu)\sigma_{\mathrm{h0}}' \tag{13}$$

As $\varepsilon_v = \varepsilon_r + \varepsilon_\theta$ under plane strain conditions, the rate form of the volumetric strain can be derived from Equation (4) as

$$\frac{\partial \varepsilon_v}{\partial t} = \frac{(1-2\mu)}{2G}\frac{\partial(\sigma_r' + \sigma_\theta')}{\partial t} \tag{14}$$

Combining Equations (7), (13), and (14), the PDE for EWP can be written as

$$\frac{\partial U}{\partial t} = c_{\mathrm{he}}\left(\frac{\partial^2 U}{\partial r^2} + \frac{1}{r}\frac{\partial U}{\partial r}\right) \tag{15}$$

$$c_{\mathrm{he}} = 2G\frac{k}{\gamma_w} \times \frac{1-\mu}{1-2\mu} \tag{16}$$

where $c_{\mathrm{he}}$ may be named as the “elastic” coefficient of consolidation.

With the boundary condition at $r = \infty$ (i.e. Equation (9)), the solution of Equation (15) can be expressed as

$$U = A_{\mathrm{e}}\mathrm{E}_1\left(r^2/4c_{\mathrm{he}}t\right) \tag{17}$$

$$\mathrm{E}_1(y) = \int_y^\infty \frac{e^{-x}}{x}\mathrm{d}x,\ y > 0 \tag{18}$$

where $A_e$ can be determined by the continuity conditions of EWP at the elastoplastic interface; $\mathrm{E}_1$ is known as the exponential integral.

Substituting Equation (13) into the equilibrium equation (3), we can get the ordinary differential equations in terms of $\sigma_r'$ and $\sigma_\theta'$ as:

$$\frac{\partial \sigma_r'}{\partial r}+\frac{2}{r}\sigma_r'=\frac{1}{r}\left[2\sigma_{\mathrm{h0}}'-\frac{U}{1-\mu}+2A_{\mathrm{e}}\exp\left(\frac{-r^2}{4c_{\mathrm{e}}t}\right)\right] \tag{19}$$

$$\sigma_\theta'=\sigma_r'+r\frac{\partial \sigma_r'}{\partial r}+r\frac{\partial U}{\partial r} \tag{20}$$

The two equations can be readily solved with the boundary conditions (9) as

$$\sigma_r'=\sigma_{\mathrm{h0}}'+B_{\mathrm{e}}\frac{\rho^2}{r^2}-\frac{U}{2(1-\mu)}+A_r \tag{21}$$

$$\sigma_\theta'=\sigma_{\mathrm{h0}}'-B_{\mathrm{e}}\frac{\rho^2}{r^2}-\frac{U}{2(1-\mu)}-A_r \tag{22}$$

$$A_r=\frac{4\mu-2}{1-\mu}\frac{A_{\mathrm{e}}c_{\mathrm{he}}t}{r^2}\exp\left(\frac{-r^2}{4c_{\mathrm{he}}t}\right) \tag{23}$$

where $B_{\mathrm{e}}$ can be determined by the yield function (5) at the elastoplastic boundary as

$$B_{\mathrm{e}}=s_{\mathrm{u}}-A_r\big|_{r=\rho} \tag{24}$$

Then substituting Equations (21) and (22) into Equation (4), the displacement is expressed as

$$\frac{u_r}{r}=\frac{1}{2G}\left[B_{\mathrm{e}}\frac{\rho^2}{r^2}+\frac{(1-2\mu)U}{2(1-\mu)}+A_r\right] \tag{25}$$

## 4. Solution in the plastic zone

This section derives the solution for the radial effective stress in the plastic zone ($a<r<\rho$) following the large strain theory.

### *4.1. Stress analysis*

Equation (6) indicates the plastic volumetric strain remains zero (i.e. $\varepsilon_r^{\mathrm{p}}+\varepsilon_\theta^{\mathrm{p}}=0$) so that $\varepsilon_v$ can be expressed as

$$\varepsilon_v=\varepsilon_r^{\mathrm{e}}+\varepsilon_\theta^{\mathrm{e}} \tag{26}$$

Then $\varepsilon_v$ can be related to $\sigma_r'$ in the rate form by substituting Equations (4) and (5) into Equation (26) as

$$\frac{\partial \varepsilon_v}{\partial t} = \frac{1-2\mu}{G}\frac{\partial \sigma_r'}{\partial t} \tag{27}$$

Combining Equations (3) and (5), the gradient of EWP can be expressed as

$$\frac{\partial U}{\partial r} = -\frac{2s_{\mathrm{u}}}{r} - \frac{\partial \sigma_r'}{\partial r} \tag{28}$$

Substituting Equations (27) and (28) into Equation (7) and eliminating $U$ and $\varepsilon_v$, the PDE in terms of $\sigma_r'$ can be derived as

$$\frac{\partial \sigma_r'}{\partial t} = c_{\mathrm{hp}}\left(\frac{\partial^2 \sigma_r'}{\partial r^2} + \frac{1}{r}\frac{\partial \sigma_r'}{\partial r}\right) \tag{29}$$

$$c_{\mathrm{hp}} = \frac{k}{\gamma_w}\frac{G}{1-2\mu} = \frac{c_{\mathrm{he}}}{2(1-\mu)} \tag{30}$$

Here $c_{\mathrm{hp}}$ may be named as the "plastic" coefficient of consolidation. Accordingly, the boundary conditions for PDE (29) can be obtained by combining Equations (10) and (28) as

$$\frac{\partial \sigma_r'}{\partial r} = \frac{-2s_{\mathrm{u}}}{a} \quad \text{for} \quad r=a<\rho \tag{31}$$

The elastoplastic stress analysis during cavity expansions under partially drained conditions was formulated as two conduction-type PDEs (i.e. Equations (15) and (29)) with moving boundary conditions at the elastoplastic interface, which is actually a typical Stefan problem (Gupta 2017). A detailed description of the solutions for the Stefan problems was summarised by Gupta (2017) with various governing PDEs and boundary conditions. Gupta (2017) demonstrated that, in the cylindrical coordinate system, exact analytical solutions for the Stefan problems exist only for some specific boundary conditions. Since $a(t)$ in Equation (31) increases continuously with time, it is hard to get a rigorous solution, if not impossible, for the present problem with the complex boundary conditions. Alternatively, it is found that an approximate solution can be derived by the variable transformation method as shown below.

For convenience, the current radius $r$ is transformed to a normalised radius $\tilde{r}$, following:

$$\tilde{r} = \frac{r}{2\sqrt{c_{\mathrm{hp}}t}} \tag{32}$$

Substituting Equation (32) into Equation (29) gives

$$\tilde{r}\frac{\partial^2 \sigma'_r}{\partial \tilde{r}^2} + \left(1 + 2\tilde{r}^2\right)\frac{\partial \sigma'_r}{\partial \tilde{r}} = 0 \tag{33}$$

Now $\sigma'_r$ that depends on *r* and *t* has been expressed as the only function of $\tilde{r}$. This transformation is applicable to the present Stefan problem and the boundary condition defined in the type of Equation (31) (Gupta 2017). If the boundary condition is shown in other forms (e.g. $\partial\sigma'_r/\partial r = \left(\sigma'_\theta - \sigma'_r\right)/a$) with more complex constitutive models, transformation of Equation (29) may involve some approximation as stated in Russell et al. (2023).

The radial effective stress can be derived from solving solving Equation (33) with boundary condition (31):

$$\sigma'_r = \sigma'_{r\rho} + s_{\mathrm{u}} e^{\tilde{a}^2}\left[\mathrm{E}_1\left(\tilde{r}^2\right) - \mathrm{E}_1\left(\tilde{\rho}^2\right)\right] \tag{34}$$

where $\sigma'_{r\rho}$ denotes the radial effective stress at $r = \rho$ and $\tilde{a}$ is the normalised inner radius. Here Equation (34) is approximate because $a(t)$ is dependent on $t$ and the boundary condition (31) cannot be rigorously transformed into a function which is only dependent on $\tilde{r}$ (Gupta 2017). However, it is later demonstrated that the error induced by this simplification is negligible by comparison with numerical simulations.

The total radial stress can be obtained by integrating Equation (2) from $\tilde{\rho}$ to $\tilde{r}$ as:

$$\sigma_r = \sigma_{r\rho} + 2s_{\mathrm{u}}\ln\left(\tilde{\rho}/\tilde{r}\right) \tag{35}$$

where $\sigma_{r\rho}$ denotes the total radial stress at $r = \rho$. Then the total radial stress at the inner cavity wall (i.e. $\sigma_a$) can be calculated from Equation (35) with $\tilde{r} = \tilde{a}$, and the EWP in the plastic zone can be obtained from Equations (34) and (35) as:

$$U = U_\rho + 2s_\mathrm{u}\ln(\tilde{\rho}/\tilde{r}) - s_\mathrm{u}e^{\tilde{a}^2}\left[\mathrm{E}_1(\tilde{r}^2) - \mathrm{E}_1(\tilde{\rho}^2)\right] \tag{36}$$

in which $U_\rho$ denotes the EWP at $r = \rho$. Determination of $\sigma'_{r\rho}$, $\sigma_{r\rho}$, and $U_\rho$ relies on the continuity conditions of stresses and EWP at the elastoplastic boundary.

### *4.2. Displacement analysis*

Logarithmic strain definitions (large srain definition) are adopted to account for the soil deformation in the plastic zone (Chadwick 1959; Yu and Houlsby 1991), and they can be expressed as Equations (37) and (38) for the present axisymmetric expansion problem.

$$\varepsilon_r = -\ln(\mathrm{d}\tilde{r}/\mathrm{d}\tilde{r}_0) \tag{37}$$

$$\varepsilon_\theta = -\ln(\tilde{r}/\tilde{r}_0) \tag{38}$$

Substituting Equations (26), (34), (37), and (38) into Equation (27) leads to

$$\ln\frac{\tilde{r}\mathrm{d}\tilde{r}}{\tilde{r}_0\mathrm{d}\tilde{r}_0} = -\omega e^{\tilde{a}^2}\left[\mathrm{E}_1(\tilde{r}^2) - \mathrm{E}_1(\tilde{\rho}^2)\right] \tag{39}$$

where $\omega = (1-2\mu)/I_\mathrm{r}$ and $I_\mathrm{r} = G/s_\mathrm{u}$ =rigidity index. Equation (39) can be further rewritten as

$$\exp\left\{\omega e^{\tilde{a}^2}\left[\mathrm{E}_1(\tilde{r}^2) - \mathrm{E}_1(\tilde{\rho}^2)\right]\right\}\mathrm{d}\tilde{r}^2 = \mathrm{d}\tilde{r}_0^2 \tag{40}$$

Integrating Equation (40) over the interval $[\tilde{a},\ \tilde{\rho}]$ gives:

$$\int_{\tilde{a}^2}^{\tilde{\rho}^2} \exp\left[\omega e^{\tilde{a}^2}\mathrm{E}_1(y)\right]\mathrm{d}y = \exp\left[\omega e^{\tilde{a}^2}\mathrm{E}_1(\tilde{\rho}^2)\right](\tilde{\rho}_0^2 - \tilde{a}_0^2) \tag{41}$$

where $\rho_0$ denotes the initial radial position of the elastoplastic interface.

Once $a(t)$ or $\tilde{a}$ is given, $\tilde{\rho}$ can be calculated from Equation (41) by simple iterations via the classical Newton-Raphson-type method. For displacement analysis, $\sigma'_{r\rho}$ in Equation (39) is simplified to be $\sigma'_{\mathrm{h}0} + s_\mathrm{u}$ (i.e. $\sigma'_{r\rho}$ does not vary with $\rho$), and thus $\rho_0$ in Equation (41) should be conjugately replaced by $\rho_0 = \rho(1-0.5/I_\mathrm{r})$ (Gibson and Anderson 1961; Yu 2000). This simplification is rigorously satisfied only for fully drained and undrained conditions (Yu 2000), but the induced error proves to be insignificant as shown later in this paper.

### *4.3 Solution at the elastoplastic interface*

The EWP as well as its gradient calculated by the elastic and plastic solutions should satisfy the continuity conditions at the elastoplastic interface. Combining Equations (17), (28), and (34), $A_e$ in Equation (17) can be derived as:

$$A_e = s_u \exp\left(\frac{\rho^2}{4c_{he}t}\right)\left(1-e^{\tilde{a}^2-\tilde{\rho}^2}\right) \tag{42}$$

Then EWP, radial effective stress, and total radial stress at $r=\rho$ can be obtained from Equations (17) and (21) yielding the following expressions:

$$U_\rho = s_u \exp\left(\frac{\rho^2}{4c_{he}t}\right)\left(1-e^{\tilde{a}^2-\tilde{\rho}^2}\right)\mathrm{E}_1\left(\frac{\rho^2}{4c_{he}t}\right) \tag{43}$$

$$\sigma'_{r\rho} = \sigma'_{h0} + s_u - \frac{1}{2(1-\mu)}U_\rho \tag{44}$$

$$\sigma_{r\rho} = \sigma_{h0} + s_u + \frac{1-2\mu}{2(1-\mu)}U_\rho \tag{45}$$

When $\tilde{\rho}$ and $\rho$ are determined from Equation (41) for a given $\tilde{a}$, the information at $r=\rho$ can be calculated by Equations (43)-(45), and finally the stresses and EWP in the plastic zone can be calculated by Equations (5) and (34)-(36). Since the present solution is shown in a straightforward form, there is no need to discretize the loading process into finite time intervals (e.g. finite increments of t or $\tilde{a}$ ), and the aforementioned limitations of the time stepping solutions can then be removed.

## 5. Special cases: ideally drained and undrained conditions

While Sections 3 and 4 give the solution for cavity expansion under partially drained condition, two special cases for ideally drained and undrained conditions can be deduced from the given solution by taking $c_p \to 0$ (ideally undrained) and $c_p \to \infty$ (ideally drained). These two cases can provide useful initial values for iterations of Equation (41) by Newton-Raphson-type methods.

### *5.1. Ideally undrained conditions*

Under undrained conditions ($c_{\mathrm{h}} \to 0$ and $\tilde{r} \to \infty$), the exponential integral can be approximately replaced by the asymptotic series:

$$e^{\tilde{a}^2}\mathrm{E}_1\left(\tilde{r}^2\right) \approx \frac{e^{\tilde{a}^2-\tilde{r}^2}}{\tilde{r}^2}\left[1+\sum_{n=1}^{10}\frac{n!}{\left(-\tilde{r}^2\right)^n}\right] < \frac{1}{\tilde{a}^2} \tag{46}$$

When $\tilde{a} \to \infty$, Equations (34) and (44) can be simplified as $\sigma'_r = \sigma'_{r\rho} = \sigma'_{\mathrm{h0}} + s_{\mathrm{u}}$ and recover Equation (41) to be:

$$a^2 - a_0^2 = \rho^2 - \rho_0^2 \tag{47}$$

Combining Equations (35) and (47), the inner cavity pressure can be expressed as

$$\sigma_a = \sigma_{\mathrm{h0}} + s_{\mathrm{u}} + s_{\mathrm{u}}\ln\left[I_{\mathrm{r}}\left(1-a_0^2/a^2\right)\right] \tag{48}$$

which is identical to the solution of Gibson and Anderson (1961). When $a_0$=0, the EWP under fully undrained conditions can be derived from Equation (48):

$$U_{\mathrm{ref}} = s_{\mathrm{u}}\ln I_{\mathrm{r}} \tag{49}$$

For the Tresca model, $U_{\mathrm{ref}}$ can be seen as the reference EWP for the normalisation of EWP in CPTU.

### *5.2. Ideally drained conditions*

Under fully drained conditions ($c_{\mathrm{p}} \to \infty$ and $\tilde{r} \to 0$), the exponential integral can be expanded into the series form as

$$\mathrm{E}_1\left(\tilde{r}^2\right) = -\zeta - 2\ln\tilde{r} - \sum_{n=1}^{\infty}\frac{\left(-\tilde{r}^2\right)^n}{nn!} \approx -\zeta - 2\ln\tilde{r} \tag{50}$$

where $\zeta$ (=0.577215…) is the Euler-Mascheroni constant. Substituting Equation (50) into Equations (34), (41), (43), and (44), we can get the analytical solution under fully drained conditions, as

$$\sigma_r' = \sigma_{h0}' + s_u + 2s_u \ln\left(\tilde{\rho}/\tilde{r}\right) \tag{51}$$

$$\left(\frac{\tilde{r}}{\tilde{\rho}}\right)^{2-2\omega} - 1 = \left(1-\omega\right)\left[\left(\frac{\tilde{r}_0}{\tilde{\rho}}\right)^2 - \left(\frac{\tilde{\rho}_0}{\tilde{\rho}}\right)^2\right] \tag{52}$$

Equations (51) and (52) are consistent with the traditional drained cavity expansion solutions that assumes no generation of EWP (Hill 1950; Yu 2000).

## 6. Results and Discussion

For illustration this section explores the cavity expansion response and the EWP evolution during cavity expansion with a constant expansion rate and a constant normalised inner radius. Unless stated otherwise, the following input parameters are used in this section: $a_0$=1 m, $u_{w0}$=0 kPa, $\sigma_{h0}$=$\sigma_{v0}$=10 kPa, $s_u$=10 kPa, $I_r$=100, $\mu$=0.3.

### *6.1. Cavity expansion with a constant expansion rate*

Cavity expansion behaviour with a constant expansion rate is firstly discussed to model the constant-rate penetration of CPTU. The inner cavity radius in this loading pattern can be defined as (Silva et al. 2006; Mafra and Dienstmann 2022):

$$a(t) = a_0 + V_a t \tag{53}$$

where $V_a$ is a (constant) cavity expansion rate. To highlight the influences of $V_a$ and relative soil permeability (i.e. $k/\gamma_w$), Figure 2 shows the cavity expansion curves and EWP at the cavity wall calculated by the present solution with $k/\gamma_w$ = $10^{-6}$ and $10^{-9}$ $m^2s^{-1}Pa^{-1}$ and various $V_a$. Cavity expansion curves under fully drained and undrained conditions are calculated by Equations (48), (51), and (52), and they serve as the lower and upper boundaries of the shadow areas in Figure 2. Numerical simulation results, taking the same soil model and input soil parameters, are computed by Comsol Multiphysics 6.0 (CM6) software with $k/\gamma_w = 10^{-6}$ $m^2s^{-1}Pa^{-1}$ and are also added for comparison.

Figure 2 illustrates that the total inner pressure under partially drained conditions is restricted within the shadow area. With the present parameters, $\sigma_a/s_\mathrm{u}$ under undrained conditions (upper bound of the shadow area) is about 4% higher than that under drained conditions (lower bound of the shadow area), which reveals that $\sigma_a/s_\mathrm{u}$ is not significantly affected by the drainage conditions and expansion rate for Tresca model. It can be found that:

(i) For cavity expansion in a high velocity in a low permeable media (e.g. $V_a = 10^{-1}$ and $10^{-9}\mathrm{m^2s^{-1}Pa^{-1}}$), the cavity behaves in an undrained manner, and the difference between $\sigma_a/s_\mathrm{u}$ and $U/s_\mathrm{u}$ is a constant ($\sigma'_{\mathrm{h0}}/s_\mathrm{u}+1$) when plasticity occurs.

(ii) For cavity expansion in a low velocity in a high permeable media (e.g. $V_a = 10^{-4}$ m/s and $10^{-6}\mathrm{m^2s^{-1}Pa^{-1}}$), the cavity shows ideally drained behaviour and $U/s_\mathrm{u}$ vanishes.

(iii) $U/s_\mathrm{u}$ at the cavity wall increases with $V_a a\gamma/k$ under partially drained conditions, indicating the significant influence of the partially drained effect. Therefore, partial drainage mainly influences the proportion of effective stresses and EWP, without significantly changing the sum of the two components (i.e. the total stress).

(iv) Comparing the EWPs in Figure 2 (a) ($k/\gamma$=$10^{-6}\mathrm{m^2s^{-1}Pa^{-1}}$) and Figure 2 (d) ($k/\gamma$ =$10^{-9}\mathrm{m^2s^{-1}Pa^{-1}}$), it is interesting to find that $U/s_\mathrm{u}$ is the same for the same $a/a_0$ and $V_a\gamma/k$. This finding will be further discussed for the normalisation of EWP under partially drained conditions.

(v) Finally, $\sigma_a/s_\mathrm{u}$ and $U/s_\mathrm{u}$ predicted by the present analytical solution agrees well with those calculated from numerical simulations, which validates the accuracy of the approximate solution.

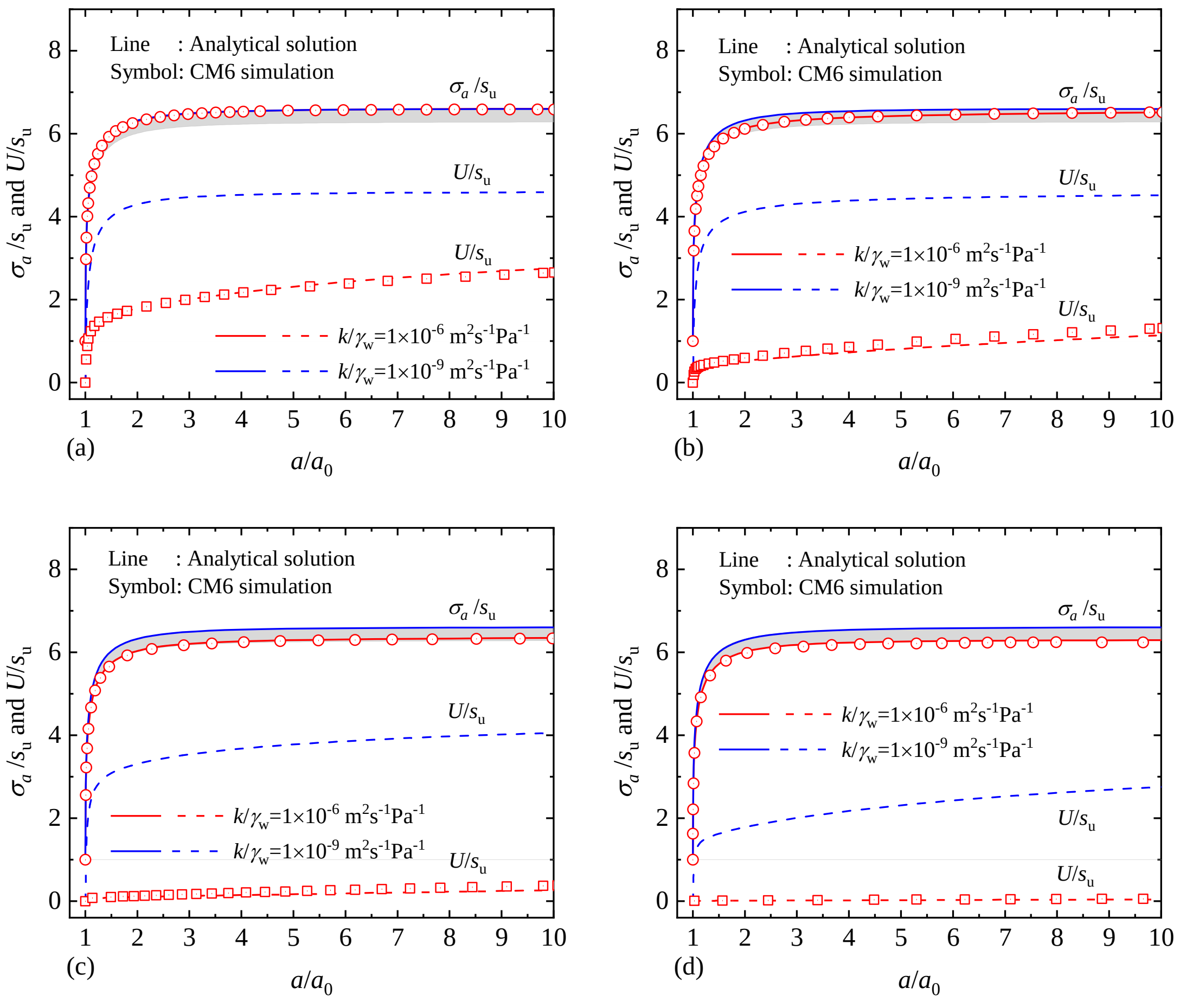

**Figure 2 Cavity expansion curves and EWP with various expansion rates:**

(a) $V_a = 10^{-1}$ m/s; (b) $V_a = 10^{-2}$ m/s; (c) $V_a = 10^{-3}$ m/s; (d) $V_a = 10^{-4}$ m/s

Alt words: The figure shows the normalised inner cavity pressure and excess water pressure change with $a/a_0$. The numerical results marked by symbols match well the the present solution marked by lines.

### *6.2. Cavity expansion with a constant normalised radius*

This subsection discusses the case of a cavity expanding with a constant normalised inner radius (i.e. constant $\tilde{a}=\delta$), in which the inner cavity radius in this loading pattern is defined as

$$a=2\delta\sqrt{c_{\mathrm{hp}}t},\ \rho \geq a \tag{54}$$

where $\delta$ is a constant that controls the "expansion rate". Substituting Equations (32) and (54) into Equation (31), the boundary conditions at $r=a$ can be simplified to

$$\frac{\partial \sigma_r'}{\partial \tilde{r}} = \frac{-2s_{\mathrm{u}}}{\delta} \quad \text{for} \quad \tilde{r}=\tilde{a}<\tilde{\rho} \tag{55}$$

In the case the right-hand side of Equation (55) is a constant and it will be proved later that the cavity expansion behaviour is independent of $k/\gamma_{\mathrm{w}}$ (or $c_{\mathrm{hp}}$ ) as long as $\delta$ is given.

The total inner pressure and EWP at the cavity wall with various $\delta$ are computed using the present solution and are plotted in Figure 3. Initially both $\sigma_a/s_{\mathrm{u}}$ and $U/s_{\mathrm{u}}$ increase with $a/a_0$ to reaching limit values for $a/a_0$ larger than 3. Due to the influence of partially drained conditions, the limit values of $\sigma_a/s_{\mathrm{u}}$ and $U/s_{\mathrm{u}}$ increase with $\delta$. The cavity expansion curves approach those under fully drained conditions for small $\delta$ (e.g. less than 0.01) and fully undrained conditions for large $\delta$ (e.g. greater than 10). Compared with the case for constant $V_a$, $U/s_{\mathrm{u}}$ computed in the case of constant $\tilde{a}$ stabilises much earlier, rather than progressively increasing with $a/a_0$ (see Figure 2). For a given value of $\delta$, the curves for constant $\tilde{a}$ are found to be independent of $k/\gamma_{\mathrm{w}}$, which represents another difference from the case for constant $V_a$. This characteristic means that $\delta$ can be back-calculated from the measured $U/s_{\mathrm{u}}$, and then $c_{\mathrm{hp}}$ may be obtained by Equation (54) with $a$ and $t$ recorded in pressuremeter tests.

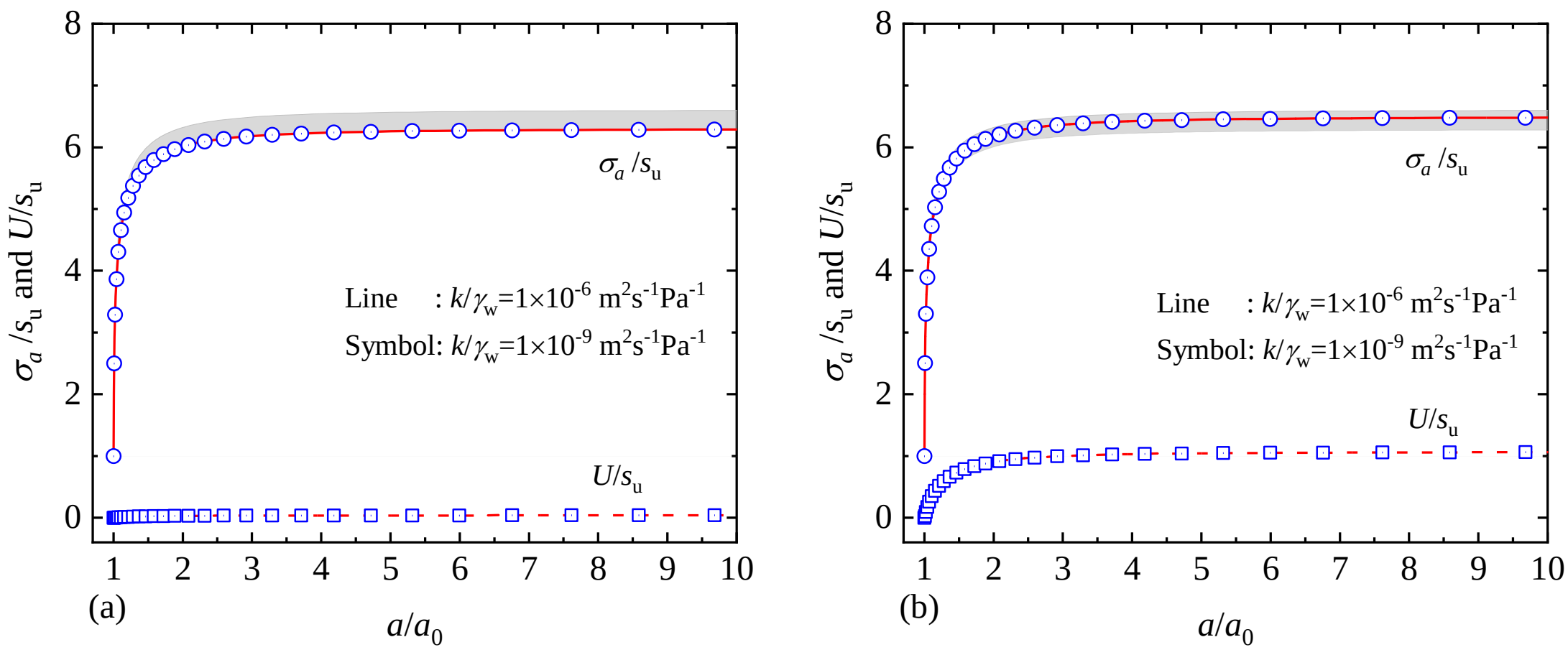

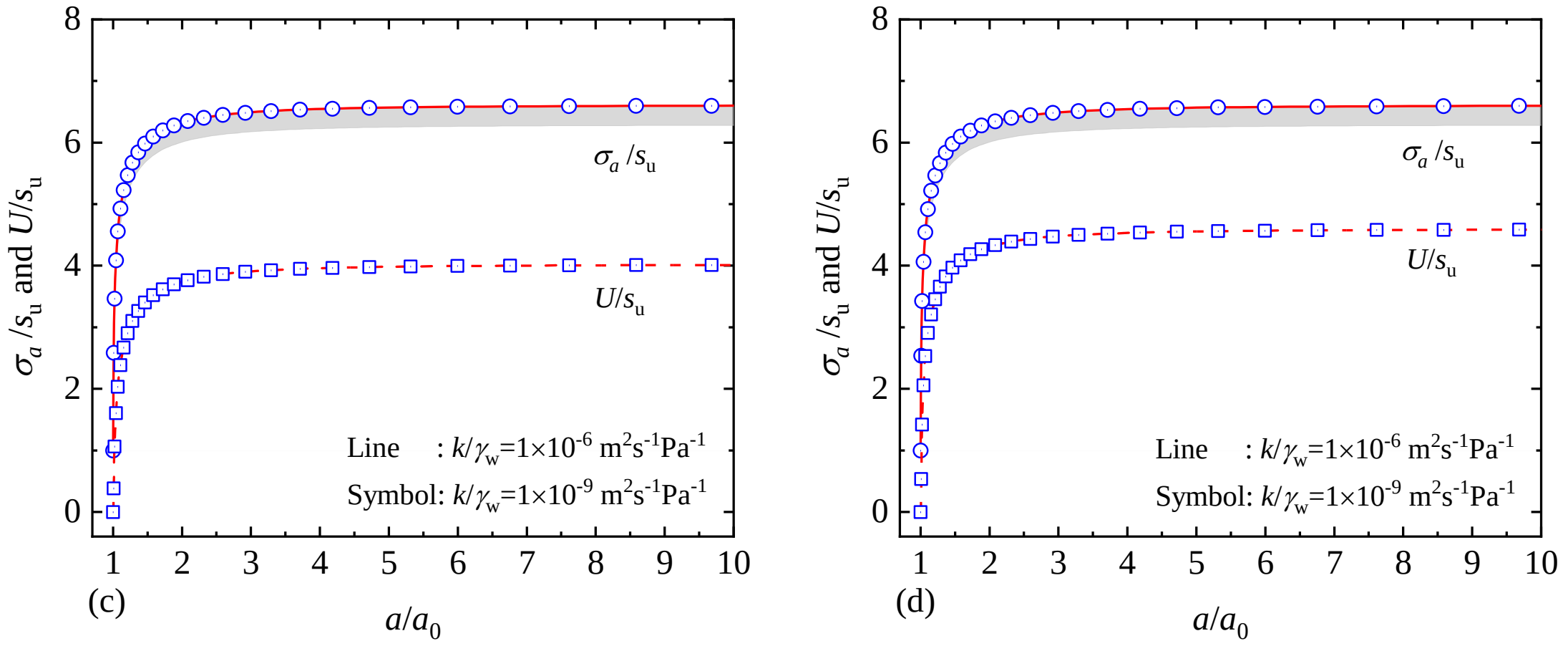

**Figure 3 Cavity expansion curves and EWP with various $\delta$:**

**(a) $\delta$ =0.01; (b) $\delta$ =0.1; (c) $\delta$ =1; (d) $\delta$ =10**

Alt words: The figure shows the normalised inner cavity pressure and excess water pressure change with $a/a_0$. As long as $\delta$ is the same, both the inner cavity pressure and excess water pressure are the same with different $k/\gamma_w$ .

## 7. Penetration rate normalisation in CPTU and potential applications

Normalisation of cone penetration rate is an important step in assessing soil properties and for comparisons of data from different sites. The normalisation of penetration rate in CPTU is discussed in this section with due consideration of partial drainage and soil rigidity variation.

### *7.1. Modified normalised rate and backbone curve*

One of the major advantages of the present straightforward-form solution lies in the much clearer physical explanation for cavity expansion under partially drained conditions. This advantage is demonstrated here by how to prove the theoretical basis of $\bar{V}_0$ for quantifying the partially drained effect in CPTU.

Combining Equations (30), (32), and (53) with $a_0 \to 0$, the normalised radius at $r = a = D/2$ can be expressed as

$$\tilde{a}^2 = \frac{V_a D}{8c_{\mathrm{hp}}} = \tan\left(\frac{\vartheta}{2}\right)\frac{\bar{V}_0}{8} \tag{56}$$

where $c_{\mathrm{hp}}$ is utilised to represent $c_{\mathrm{h}}$ in Equation (1), which enables $\bar{V}_0$ to be expressed as a function of $\tilde{a}^2$ only ($\vartheta$ is usually $60^\circ$). Once $\tilde{a}^2$ (or $\bar{V}_0$) and $I_{\mathrm{r}}$ are given, $\tilde{\rho}^2$ can be fully determined from Equation (41). Then the normalised EWP, $\bar{U}$ (see Equation (57)), can also be known by Equations (36), (43), and (49).

$$\bar{U} = U/U_{\mathrm{ref}} \tag{57}$$

Therefore, the backbone curve (i.e. $\bar{U}-\bar{V}_0$ curve) is unique for a given soil (i.e. given $I_{\mathrm{r}}$), and $\bar{V}_0$ can be used to account for the partial drainage conditions.

Equations (41) and some former publications (Dienstmann et al. 2018; Mafra and Dienstmann 2022) show that the evolution of EWP is also affected by $I_{\mathrm{r}}$. When $\bar{U}$ =0.5, for example, Dienstmann et al. (2018) reported that $\bar{V}_0 \approx 0.01$ for $I_{\mathrm{r}}=900$ while $\bar{V}_0 \approx 0.03$ for $I_{\mathrm{r}}=100$. Therefore, $\bar{V}_0$ may not fully quantify the partial drainage effect for a wide range of soils with various $I_{\mathrm{r}}$. Moreover, Figure 4 (a) plots 6 series of $\bar{U}-\bar{V}_0$ curves calculated by the present solution with input parameters in Table 1. It is clearly shown that, under partially drained conditions, $\bar{U}$ increases with $I_{\mathrm{r}}$ for a given $\bar{V}_0$, and the backbone curves are noticeably dependent on $I_{\mathrm{r}}$.

To consider the influence of $I_{\mathrm{r}}$ in the penetration process, a modified normalised penetration rate, $\bar{V}$, is defined in this paper as

$$\bar{V} = \bar{V}_0 I_{\mathrm{r}}^m = \frac{V_{\mathrm{cptu}} D}{c_{\mathrm{hp}}} I_{\mathrm{r}}^m \tag{58}$$

where *m* is a curve fitting parameter in the range of 0.45-0.5. The calculated backbone curves are replotted in Figure 4 (b) with the modified $\bar{V}$ (*m*=0.5), and it is shown that the modified backbone curves ($\bar{U}-\bar{V}$) converge to a narrow band that can be fitted by the following expression (DeJong and Randolph 2012):

$$\bar{U} = 1 - \frac{1}{1 + \left(\bar{V}/8.1\right)^{0.7}} \tag{59}$$

The uniqueness of Equations (58) and (59) for normalisation of partially drained effect is clearly illustrated.

**Table 1. Input parameters for the normalisation of rate effect**

| Series | $G$ : kPa | $s_u$ : kPa | $I_r$ | $U_{ref}/s_u$ |
|---|---|---|---|---|
| S1 | 200 | 10 | 20 | 3.00 |
| S2 | 500 | 10 | 50 | 3.91 |
| S3 | 1000 | 10 | 100 | 4.60 |
| S4 | 2000 | 10 | 200 | 5.29 |
| S5 | 5000 | 10 | 500 | 6.20 |
| S6 | 10000 | 10 | 1000 | 6.90 |

Note: $V_a = 10^{-2}$ m/s, $k/\gamma_w = 10^{-5} \sim 10^{-15}$ m$^2$s$^{-1}$Pa$^{-1}$, $\mu$=0.3, $a_0$=0, and $D$=0.04m. These input parameters will not affect the main findings after dimensionless normalisation.

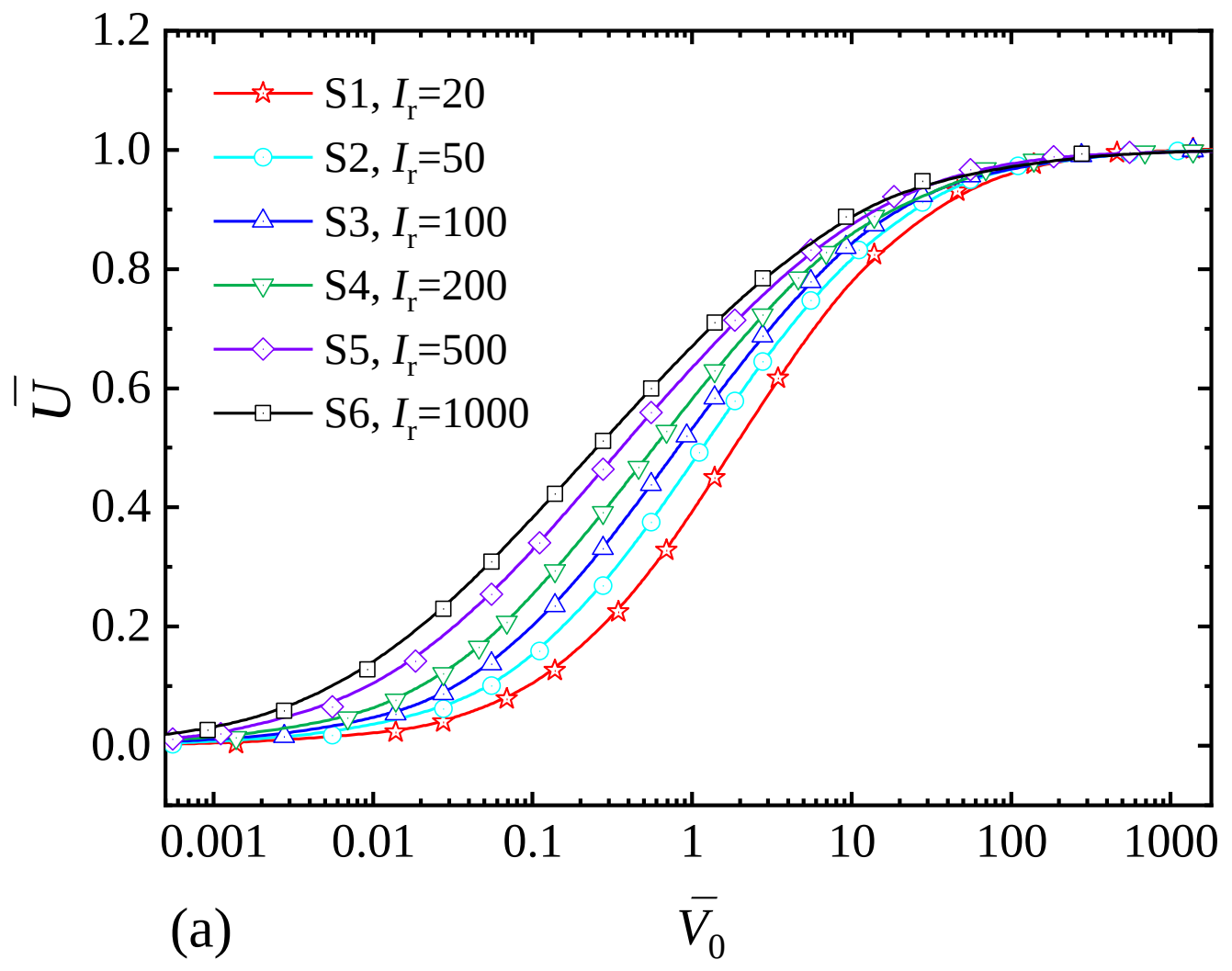

(a)

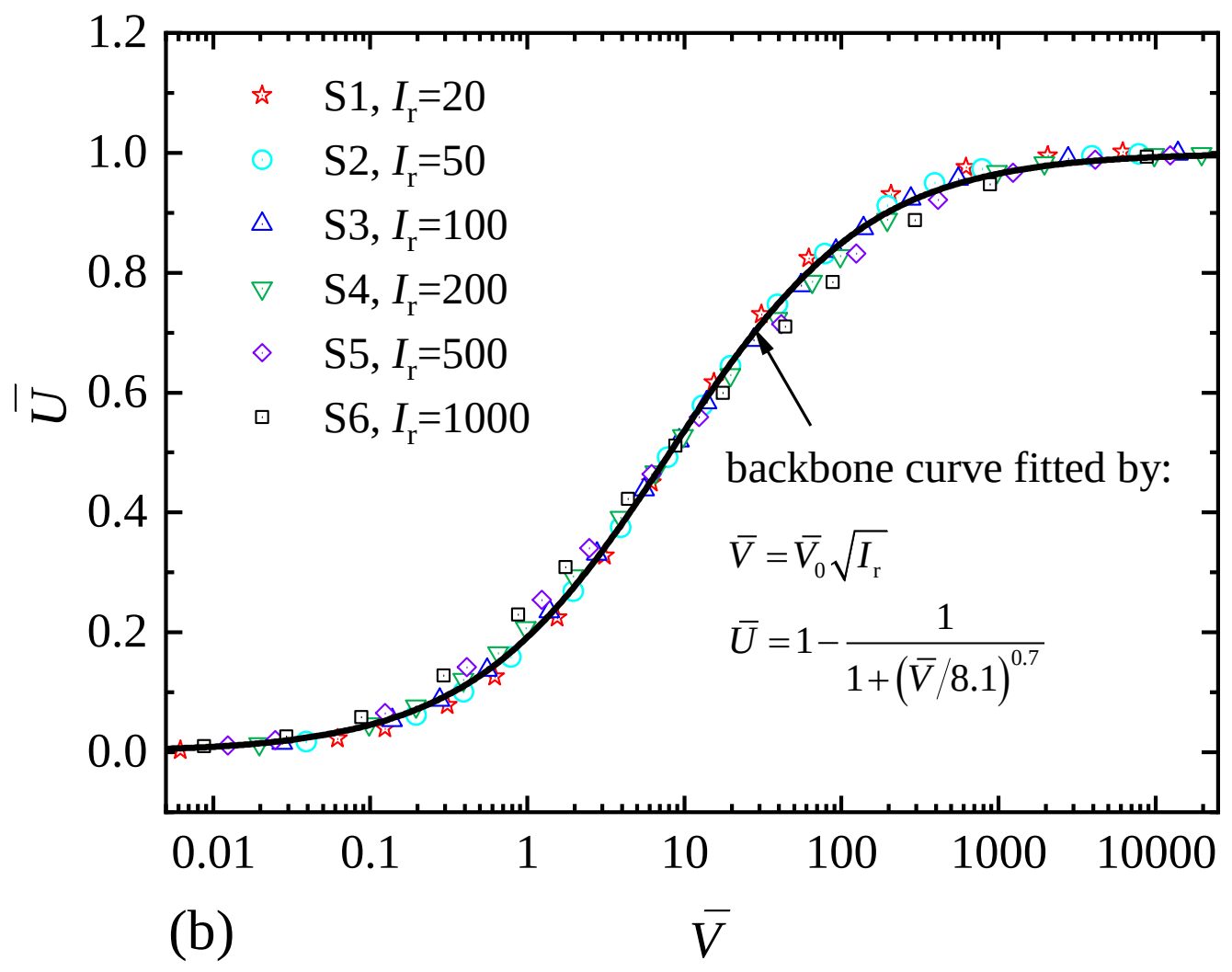

**Figure 4 Backbone curves in CPTU: (a) based on $\bar{V}_0$; (b) based on $\bar{V}$**

Alt words: In figure 4a, the backbone curves are different with different $I_r$, and the backbone curve on the right part for a small $I_r$. After normalisation by a new nomalised penetration rate, these backbone curves are restricted into a very narrow band.

### *7.2. Validation of the backbone curve and potential applications*

The new backbone curve defined in Equation (59) is validated by comparison with a database containing experimental and numerical results, as summarised in Table 2. In the database a total number of 109 in-situ tests on silts-gold tailings (Fazenda Brasileiro Mine, Brazil) and 101 Centrifuge model tests (40g) on normally consolidated UWA Kaolin Clay are collated. Numerical simulation results were mainly calculated by FEM and MPM, and the Drucker-Prager model (DP) and MCC were used for constitutive modelling of soils (e.g. Kaolin clays and Malaysian Kaolin silt). Overall, the database covers a wide range of $I_r$ from 32 to 874.

The database data are plotted in Figure 5 for comparison with the proposed solution (see Equation (59)). The scatter of the database is mainly from in-situ tests on silts-gold tailings, which may be attributed to the wide spatial dispersion of testing sites, deposition conditions, material segregation, and other factors (Dienstmann et al. 2018; Mafra and Dienstmann 2022). When compared with the database, the new backbone curve predicted by the present solution is able to capture well the partially drained effect in CPTU. Besides, the present backbone curve gives an overestimation of $\bar{U}$ for a small $\bar{V}$ (i.e. approximately less than 3), and it may be induced by the neglect of EWP dissipation and soil deformation in the vertical direction (Soderberg 1962; Robertson and Campanella 1983a; Zhang et al. 2022). The vertical EWP dissipation may be partially described by a corresponding solution in the spherical scenario, and this will be developed in a further study combining with the normalisation of cone tip resistance. Figure 5 also compares the backbone curves predicted by Mafra and Dienstmann (2022) and this study via the cavity expansion method. The two curves show similar trends but the curve of Mafra and Dienstmann (2022) tends to overestimate $\bar{U}$. This difference may result from the imposed initial EWP distribution predefined in Mafra and Dienstmann (2022), which is not required in the proposed solution.

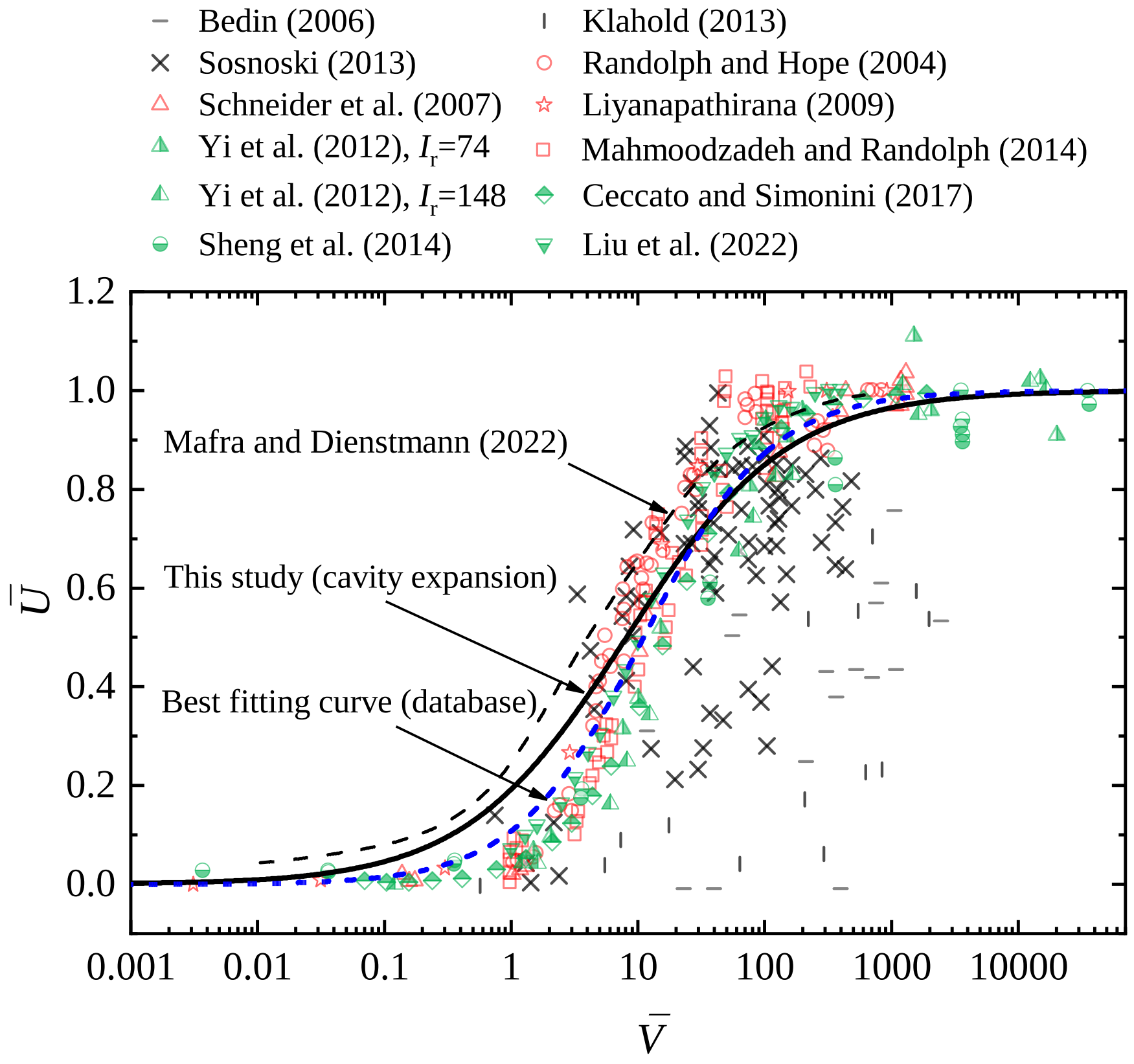

Figure 5 Predicted backbone curve versus experimental and numerical results

Alt words: In this figure the data from the database show a general trend that U increases with V and becomes 1 for a very large V (e.g., V>2000). The backbone curves using cavity expansion theory can generally capture this trend. For a same U, the backbone curve in this study predicts a higher V than the curve of Mafra and Dienstmann (2022), and matches better with the database.

**Table 2. Database and cavity expansion method for partially drained effect in CPTU**

| References | Soil type/ model | Numbers | $I_r$ | $c_{he}/c_v$ | $\mu$ | Method |
|---|---|---|---|---|---|---|
| Bedin (2006) | Silty-gold tailing | 16 | 874[&] | — | 0.3 | In-situ tests |
| Klahold (2013) | Silty-gold tailing | 14 | 874[&] | — | 0.3 | |
| Sosnoski (2016) | Silty-gold tailing | 79 | 874[&] | — | 0.3 | |
| Randolph and Hope (2004) | Kaolin clay | 50 | 76* | 4.66* | 0.3 | Centrifuge tests |
| Schneider et al. (2007) | Kaolin clay | 21 | 76* | 4.66* | 0.3 | |
| Mahmoodzadeh and Randolph (2014) | Kaolin clay | 30 | 76* | 4.66* | 0.3 | |
| Liyanapathirana (2009) | MCC | 9 | 74 | 4.66* | 0.3 | FEM |
| Yi et al. (2012) | DP | 40 | 74/148 | 1 | 0.3 | |
| Sheng et al. (2014) | MCC | 23 | 32[§] | — | 0.33 | |
| Liu et al. (2022) | MCC | 24 | 59[†] | 4.7[‡] | 0.3 | |
| Ceccato and Simonini (2017) | MCC | — | 108 | 3 | 0.25 | MPM |
| Mafra and Dienstmann (2022)** | DP | — | 874 | — | 0.3 | Cavity expansion |
| This study | Tresca | — | | — | 0.3 | Cavity expansion |

[&]Refer to Dienstmann et al. (2017), Dienstmann et al. (2018), and Schnaid et al. (2020)

*Refer to Mahmoodzadeh and Randolph (2014) and Mahmoodzadeh et al. (2015)

[§]Refer to Obrzud et al. (2011)

[†]Refer to Zhang et al. (2022)

[‡]Variation of $c_{he}/c_v$ with $\bar{V}_0$ is not considered

**The curve without correction is selected (i.e. $I_r \approx 874$)

The unique backbone curve ($\bar{U}-\bar{V}$) may offer a potential approach to assess the in-situ consolidation coefficient of soils from multi-rate CPTU (Fahey and Goh 1995; House et al. 2001; Randolph and Hope 2004), following:

(i) Carry out multi-rate CPTU with different penetration rates and measure EWPs.

(ii) Interpret $I_{\mathrm{r}}$ by conventional methods (e.g. Mayne et al. (2022); Khodayari and Ahmadi (2022)).

(iii) Combine Equations (58) and (59) to calculate $c_{\mathrm{hp}}$ by Equation (60) with $I_{\mathrm{r}}$ and the measured EWPs.

$$c_{\mathrm{hp}} = \frac{V_{\mathrm{cptu}} D \sqrt{I_{\mathrm{r}}}}{8.1} \left( \frac{1-\bar{U}}{\bar{U}} \right)^{1.43} \tag{60}$$

In the future multi-rate CPTU and subsequent dissipation tests can be performed both in the field and in the laboratory, and it is expected that more reliable values of $c_{\mathrm{h}}$ can be obtained by combining the interpretation of CPTU in the penetration and dissipation stages. Finally, it is worth noticing that although the partially drained effect in CPTU is only quantified here by the normalised EWP, due to the limitation of paper length, further research will focus on the normalisation of cone tip resistance by $\bar{V}$.

## 8. Discussion on the selected Tresca model

The major limitation of the present solution is the selection of the Tresca model to develop the hydro-mechanical coupled solution for cavity expansion under partially drained conditions. The inherent shortcomings of the Tresca model are therefore inclined in the solution:

(i) The equivalent shear strength and shear modulus are input parameters rather than being dependent on stress and volume changes. Therefore, the state-dependent soil strength and stiffness are beyond the scope of this paper.

(ii) The soil model fails to encompass the consequence of volumetric change on the soil mechanical behaviour because the plastic volumetric strain is always zero with the associated plastic flow rule (i.e. Equation (6)). Accordingly, parameters controlling the compressibility of a normally consolidated clay have not been included in the calculation of "plastic" coefficient of consolidation.

However, an elastoplastic, hydro-mechanical coupled solution for cavity expansion has not been reported so far in a straightforward form, even when using such a simple Tresca model. The advantages of the present solution lie in that:

(i) From the mathematical point of view, the Tresca mode enables the cavity expansion analysis to be formulated into a standard Stefan problem and solved by the variable transformation method. When more advanced soil models are adopted (e.g. MCC in the time-stepping solution of Russell et al. (2023)), it may not be suitable to solve the complex governing equations by the variable transformation method.

(ii) From the perspective of solution form, choosing the Tresca model can make the solution be expressed in a concise straightforward form. Only in this concise form the normalisation of EWP can be deeply understood with the strength of the solution in dimensional analysis.

(iii) From the practical point of view, there is an urgent need to modify the original normalised penetration rate for CPTU by considering $I_{\mathrm{r}}$. Since $I_{\mathrm{r}}$ is the function of $s_{\mathrm{u}}$ and *G*, it can be more convenient to pay attention to these two parameters, and Tresca model will be a valuable choice. The validation of the new normalised penetration rate and backbone are also helpful to demonstrate the reasonability of this simplification.

(iv) In view of the comparison of different soil models, the backbone curves behave in a similar trend as shown in Figure 5. It may be explained by: the generation and dissipation of the normalised EWP are mainly controlled by Darcy's law that is adopted in the database with different soil models. Therefore, this may be the reason why the present solution with the simple Tresca model can predict the new backbone curve successfully.

In summary, this paper contributes to the development of the cavity expansion theory under partially drained conditions and its application in geotechnical engineering, with the Tresca model selected as a starting point.

## 9. Conclusions

This paper proposes a hydro-mechanical coupling solution for cylindrical cavity expansion under partially drained conditions. The soil is modelled as a perfectly elastoplastic Tresca material and the hydraulic behaviour is assumed to obey Darcy's law. Two PDEs are formulated for the cavity expansion analysis in the elastic and plastic zones, respectively, and they are solved by the variable transformation method. The new solution is initially validated by comparison with numerical simulation results. Subsequent parametric studies show that the partially drained effect has an important influence on EWP evolution but little on the total inner pressure.

The proposed solution is applied to the normalisation of EWP in CPTU. The physical meaning and limitations of the existing normalised penetration rate are discussed to highlight the advantages of the present solution, and a modified normalised penetration rate is defined by considering the influence of the rigidity index of soils. With the new normalised rate, a unique backbone curve is found and validated against a database of experimental tests and numerical simulation results. Finally, a method for interpreting multi-rate CPTU (penetration stage) is suggested to measure the coefficient of consolidation of soils. Overall the paper contributes to quantifying the partially drained effect in CPTU to enhance the interpretation of CPTU in future applications.

## Data Availability Statement

Data are available from the corresponding author upon reasonable request.

## Acknowledgement

The first author thanks the financial support from the China Scholarship Council for his study at the University of Leeds. The second author would like to acknowledge the financial support from the National Natural Science Foundation of China (52108374), the "Taishan" Scholar Program of Shandong Province, China (tsqn201909016), the Shandong Provincial Natural Science Foundation (ZR202102250562), and the "Qilu" Young Scholar Program of Shandong University.

**Competing Interests**

The authors declare that there is no known competing interests.

**Notation List**

| | |
|---|---|
| $a_0$, $a$ | initial and current radii of the inner cavity wall |
| $\tilde{a}$ | normalised inner radius at the inner cavity wall |
| $A_\mathrm{e}$ | integral constant for excess water pressure |
| $c_\mathrm{h}$, $c_\mathrm{v}$ | horizontal and vertical coefficients of consolidation |
| $c_\mathrm{he}$, $c_\mathrm{hp}$ | elastic and plastic coefficients of consolidation |
| $D$ | piezocone diameter |
| $\mathrm{E}_1$ | exponential integral |
| $G$ | shear modulus of soils |
| $I_\mathrm{r}$ | rigidity index of soils |
| $k$ | permeability coefficient of soils |
| $m$ | curve fitting parameter for backbone curve |
| $r_0$, $r$ | initial and current radial positions of a soil particle |
| $\tilde{r}$ | normalised radial position of a soil particle |
| $s_\mathrm{u}$ | shear strength of soils |
| $t$ | time and $t$=0 for in-situ stress state |
| $u_r$ | radial displacement of a soil particle |
| $u_\mathrm{w0}$ | initial ambient water pressure |
| $U$ | excess water pressure |
| $U_\rho$ | excess water pressure at the elastoplastic boundary |
| $U_\mathrm{ref}$, $\bar{U}$ | reference and normalised excess water pressures |
| $V_a$, $V_\mathrm{cptu}$ | rates of cavity expansion and penetration |
| $\bar{V}_0$, $\bar{V}$ | normalised penetration rates with and without modifications |
| $v$ | specific volume of soils |
| $\delta$ | constant controlling cavity expansion rate |
| $\varepsilon_r^{\mathrm{e}}$, $\varepsilon_\theta^{\mathrm{e}}$ | elastic radial and circumferential strains |
| $\varepsilon_r^{\mathrm{p}}$, $\varepsilon_\theta^{\mathrm{p}}$ | plastic radial and circumferential strains |
| $\varepsilon_r$, $\varepsilon_\theta$, $\varepsilon_{\mathrm{v}}$ | total radial, circumferential, and volumetric strains |
| $\gamma_{\mathrm{w}}$ | specific gravity of soils |
| $\mu$ | Poisson's ratio of soils |
| $\omega$ | soil parameter dependent on Poisson's ratio and rigidity index |
| $\rho_0$, $\rho$ | initial and current radii of the elastoplastic interface |
| $\theta$ | piezocone tip angle |
| $\sigma_a$ | inner cavity pressure |
| $\sigma_{\mathrm{h0}}$, $\sigma_{\mathrm{v0}}$ | initial horizontal and vertical total stresses |
| $\sigma'_{\mathrm{h0}}$, $\sigma'_{\mathrm{v0}}$ | initial horizontal and vertical effective stresses |
| $\sigma_r$, $\sigma_\theta$ | total radial and circumferential stresses |
| $\sigma'_r$, $\sigma'_\theta$ | effective radial and circumferential stresses |
| $\sigma_{r\rho}$, $\sigma'_{r\rho}$ | total and effective radial stresses at the elastoplastic interface |